\documentstyle[12pt]{article}
\input epsf

\newcommand{\bq}{\begin{equation}}
\newcommand{\eq}{\end{equation}}
\newcommand{\bqa}{\begin{eqnarray}}
\newcommand{\eqa}{\end{eqnarray}}

\newcommand{\ff}{{\ifmmode D(x,Q^2)    \else $D(x,Q^2)$\fi}}
\newcommand{\lnf}{{\ifmmode \Lambda^{(N_f)} \else $\Lambda^{(N_f)}$\fi}}
\newcommand{\ms}{{\ifmmode \overline{MS} \else $\overline{MS}$\fi}}
\newcommand{\lms}{{\ifmmode \Lambda^{(5)}_{\overline{MS}}
                 \else $\Lambda^{(5)}_{\overline{MS}}$\fi}}
\newcommand{\lam}{{\ifmmode \Lambda \else $\Lambda$\fi}}
\newcommand{\gev}{{\ifmmode {\rm GeV/c^2} \else ${\rm GeV/c^2}$\fi}}
\newcommand{\lp}{{\ifmmode L^+  \else $L^+$\fi}}
\newcommand{\lm}{{\ifmmode L^-  \else $L^-$\fi}}
\newcommand{\mlp}{{\ifmmode M(L^-)  \else $M(L^-)$\fi}}
\newcommand{\mlz}{{\ifmmode M(L^0)  \else $M(L^0)$\fi}}
\newcommand{\lz}{{\ifmmode L^0     \else $L^0$\fi}}
\newcommand{\ev}{{\ifmmode GeV/c^2       else $GeV/c^2$\fi}}
\newcommand{\tri}{{\ifmmode \triangleup  \else $\triangleup$\fi}}
\newcommand{\unl}{{\ifmmode U_{lL^0}  \else $U_{lL^0}$\fi}}
\newcommand{\gL}{{\ifmmode g_L  \else $g_{L}$\fi}}
\newcommand{\gR}{{\ifmmode g_R  \else $g_{R}$\fi}}
\newcommand{\gumu}{{\ifmmode \gamma^{\mu}  \else $\gamma^{\mu}$\fi}}
\newcommand{\gunu}{{\ifmmode \gamma^{\nu}  \else $\gamma^{\nu}$\fi}}
\newcommand{\gdmu}{{\ifmmode \gamma_{\mu}  \else $\gamma_{\mu}$\fi}}
\newcommand{\gdnu}{{\ifmmode \gamma_{\nu}  \else $\gamma_{\nu}$\fi}}
\newcommand{\stw}{{\ifmmode\sin^2\theta_W  \else $\sin^{2}\theta_{W}$
\fi}}
\newcommand{\sw}{{\ifmmode \sin^2\theta_W  \else $\sin^{2}\theta_{W}$
\fi}}
\newcommand{\swb}{{\ifmmode \sin^2\theta_{\overline{MS}}
                     \else $\sin^2\theta_{\overline{MS}}$
\fi}}
\newcommand{\cwb}{{\ifmmode \cos^2\theta_{\overline{MS}}
                     \else $\cos^2\theta_{\overline{MS}}$
\fi}}
\newcommand{\qq}{{\ifmmode q\overline{q} \else $q\overline{q}$\fi}}
\newcommand{\as}{{\ifmmode \alpha_s  \else $\alpha_s$\fi}}
\newcommand{\lR}{{\ifmmode l_R  \else $l_R$\fi}}
\newcommand{\lL}{{\ifmmode l_L  \else &l_L$\fi}}
\newcommand{\pperp}{{\ifmmode p_t  \else $p_t$\fi}}
\newcommand{\et}{{\ifmmode E_t  \else $E_t$\fi}}
\newcommand{\xt}{{\ifmmode x_t  \else $x_t$\fi}}
\newcommand{\smumu}{{\ifmmode \sigma_{\mu\mu}  \else $\sigma_{\mu\mu}$
\fi}}
\newcommand{\eg}{{\ifmmode e\gamma  \else $e\gamma$\fi}}
\newcommand{\epem}{{\ifmmode e^+e^-  \else $e^+e^-$\fi}}
\newcommand{\lplm}{{\ifmmode L^+L^-  \else $L^+L^-$\fi}}
\newcommand{\pp}{{\ifmmode p\overline p  \else $p\overline p$\fi}}
\newcommand{\llz}{{\ifmmode L^0\overline{L}^0 \else
$L^0\overline{L}^0$\fi}}
\newcommand{\epemt}{{\ifmmode e^+e^- \to  \else $e^+e^- \to$\fi}}
\newcommand{\eb}{{\ifmmode E_{beam}  \else $E_{beam}$\fi}}
\newcommand{\ip}{{\ifmmode pb^{-1}  \else $pb^{-1}$\fi}}
\newcommand{\upm}{{\ifmmode ^{\pm}  \else $^{\pm}$\fi}}
\newcommand{\de}{{\ifmmode ^{\circ}  \else $^{\circ}$ \fi}}
\newcommand{\appr}{{\ifmmode \sim \else $\sim$ \fi}}
\newcommand{\corresp}{{\ifmmode \stackrel{\wedge}{=}
                      \else   $\stackrel{\wedge}{=}$ \fi}}
\newcommand{\sqrts}{{\ifmmode \sqrt{s} \else $\sqrt{s}$\fi}}
\newcommand{\zz}{{\ifmmode Z^0  \else $Z^0$\fi}}
\newcommand{\mz}{{\ifmmode M_{Z}  \else $M_{Z}$\fi}}
\newcommand{\mw}{{\ifmmode M_{W}  \else $M_{W}$\fi}}
\newcommand{\mh}{{\ifmmode M_{Higgs}  \else $M_{Higgs}$\fi}}
\newcommand{\gt}{{\ifmmode \Gamma_{tot} \else $\Gamma_{tot}$\fi}}
\newcommand{\mt}{{\ifmmode M_{top}  \else $M_{top}$\fi}}
\newcommand{\msusy}{{\ifmmode M_{SUSY}  \else $M_{SUSY}$\fi}}
\newcommand{\taup} {{\ifmmode \tau_{proton} \else $\tau_{proton}$\fi}}
\newcommand{\mgut}{{\ifmmode M_{GUT}  \else $M_{GUT}$\fi}}
\begin{document}


{\flushright\hfill \bf\normalsize IEKP--KA/93--8\\

                        6th July 1993\\
      Ext. Version of  CERN--PPE/93--69\\}
\vspace*{1.0cm}

\begin{center}
\LARGE
 Scaling Violation \\in the Fragmentation Functions \\
in  $e^{+}e^{-}$-Annihilation

\vspace*{0.7cm}
{\large
W. de Boer and T. Ku\ss maul\\}
\vspace*{0.3cm}

{\normalsize
Institut f\"ur Experimentelle Kernphysik\\
Universit\"at Karlsruhe, Postfach 6980,\\
D-76128 Karlsruhe\\}


\end{center}


\vspace*{4.5cm}
\begin{abstract}
A determination
                 of the hadronic fragmentation functions of the
$\zz$ boson is presented from a study of the inclusive
hadron production wit the DELPHI detector at LEP.

These fragmentation functions were      compared with the
ones at lower energies, thus covering data in a large kinematic
range:
  $196\leq Q^2 \leq 8312 ~ GeV^2$ and \mbox{$x(= p_h/E_{beam}) > 0.08$.}
A large scaling violation was observed, which was
                      used to extract the strong coupling
constant in second order QCD:
                   $$\as(\mz) =0.118\pm 0.005. $$
The  corresponding
                 QCD scale for five quark flavours is:
 $\lms
=230 \pm 60$              MeV.

\begin{center}
\vglue 0.8cm
(Contribution to the XVI International Symposium on \\
Lepton-Photon Interactions, \\
 Cornell, 10-15 August,
1993)
\end{center}
\end{abstract}

\newpage
\section{Introduction}
Hadron production in $ e^+e^-$ annihilation
originates from the production of quark-antiquark pairs,
which can radiate gluons, the  quanta of the  field theory
of the strong interactions, Quantum ChromoDynamics (QCD).
     Gluon radiation depends logarithmically
on the centre of mass energy due to the
increasing phase  space with increasing energy and
the energy dependence of the running coupling constant of QCD.
These effects lead to variations of the
momentum   spectra of the produced hadrons  as a function
of the centre of mass energy, even if the momenta are scaled to
that energy.   These scaling violations can be used to
determine the strong coupling constant \as.

     For example,
the  scaling violation in deep
inelastic lepton--nucleon scattering
leads to
    $\as(\mz)=0.112\pm0.005 $ \cite{vir,mar,qui}.
     This is  somewhat lower than, but
not in disagreement
with, $\as$ measurements at the $\zz$ mass from
shape  variables, jet rates and total cross sections
as measured
 at the electron-positron storage ring LEP\cite{lep,rev}.

  Until  now \as\ has not been determined from scaling
 violations   in  $\epem$
annihilation,   since this requires
 precise data at  two very different energies  to observe a
 significant scaling violation.
Data collected at the PEP and PETRA storage rings were
 only precise  at energies
around  $\sqrts=30$ GeV\cite{tas,cel,mk2}
and the scaling violation was only observed qualitatively\cite{tas0}.

In this paper we present data of the inclusive hadron
spectra, as measured with the DELPHI detector\cite{det} at LEP
and present the first \as\ determination from the scaling
violations in the fragmentation function  by combining the
data at LEP with data from the
PEP, PETRA and TRISTAN storage rings. 
The squared four-momentum transfer
from the incoming leptons to  outgoing
hadrons studied here
is two orders of magnitude larger than the ones
studied in deep inelastic scattering,  thus   avoiding
regions where non-perturbative effects noticeably influence the results.
On the other hand, in $\epem$ annihilation one has to combine
data from different experiments at different energies
and study the effect of the varying quark flavour composition
due to the $\zz$-resonance. As will be shown, these are
not  dominant uncertainties (see also Ref.      \cite{wdb}).

\section{Determination of the Fragmentation Function }
   The inclusive  production of charged hadrons  in the reaction
$ e^{+}e^{-} \rightarrow h + X $ can be
described by          two kinematic
variables, $Q^2$ and $x$, where $Q^2$ is defined as the
       square of the four-momentum transferred from the leptons
  to the hadrons
and $x$ is the
fraction     of the beam energy $ E_{beam} $ carried by the hadron $h$.
In $\epem$ annihilation $Q^2$ equals $s$,
 the total centre of mass energy squared.

  The fragmentation function  $\bar{D}(x,Q^2)$
  is  directly related to the scaled
  hadron momentum distribution:
  \begin{equation}
  \bar{D}(x,Q^2) \equiv \sum  \limits_{i=1}^{5}
   W_i(Q^2){    {D}_i(x,Q^2)}
     \equiv \frac{1}{\sigma_{t}} \frac{d\sigma}{dx} (\epemt h+X),
     \label{e1}
  \end{equation}
  where $\sigma_t$ is the total cross section and
                  $\bar{D}(x,Q^2)$  is the sum of fragmentation
  functions $    {D}_i(x,Q^2)$ over all five flavours,
   each having a weight $W_i(Q^2)$
  given by the electroweak theory.

The DELPHI data were      collected during 1991  at energies near the
\zz\ peak.
Multi-hadronic events
were selected according to the criteria given in
Ref.\cite{STDRCUT}. The selection required that there were at least
5 charged particles
 with momenta above 0.2 GeV/c and a track length in the detector of
at least 50 cm,
that the   total  energy  of the
charged particles          exceeded 15 GeV (pion
mass assumed), that the forward    and backward
 hemisphere with respect to the beam axis
each contained a total  energy
of charged particles   larger than 3 GeV, and that the polar
 angle of the sphericity axis was between
$40^o$ and $140^o$.
In addition the momentum imbalance was restricted by
requiring that         the absolute sum of the three-momenta
    of charged particles was less than  20 GeV/c.

After these selection criteria,
      186774 events at a mean centre of mass
energy of 91.2 GeV were kept.
The background due to beam-gas scattering and
$\gamma \gamma$-interactions was less than 0.1\% and $\tau^+ \tau^-$
events contributed    0.2\% to the selected sample.

\begin{figure}[b]
\vspace*{1.0cm}

\begin{center}
\mbox{\epsfysize=20cm\epsfbox{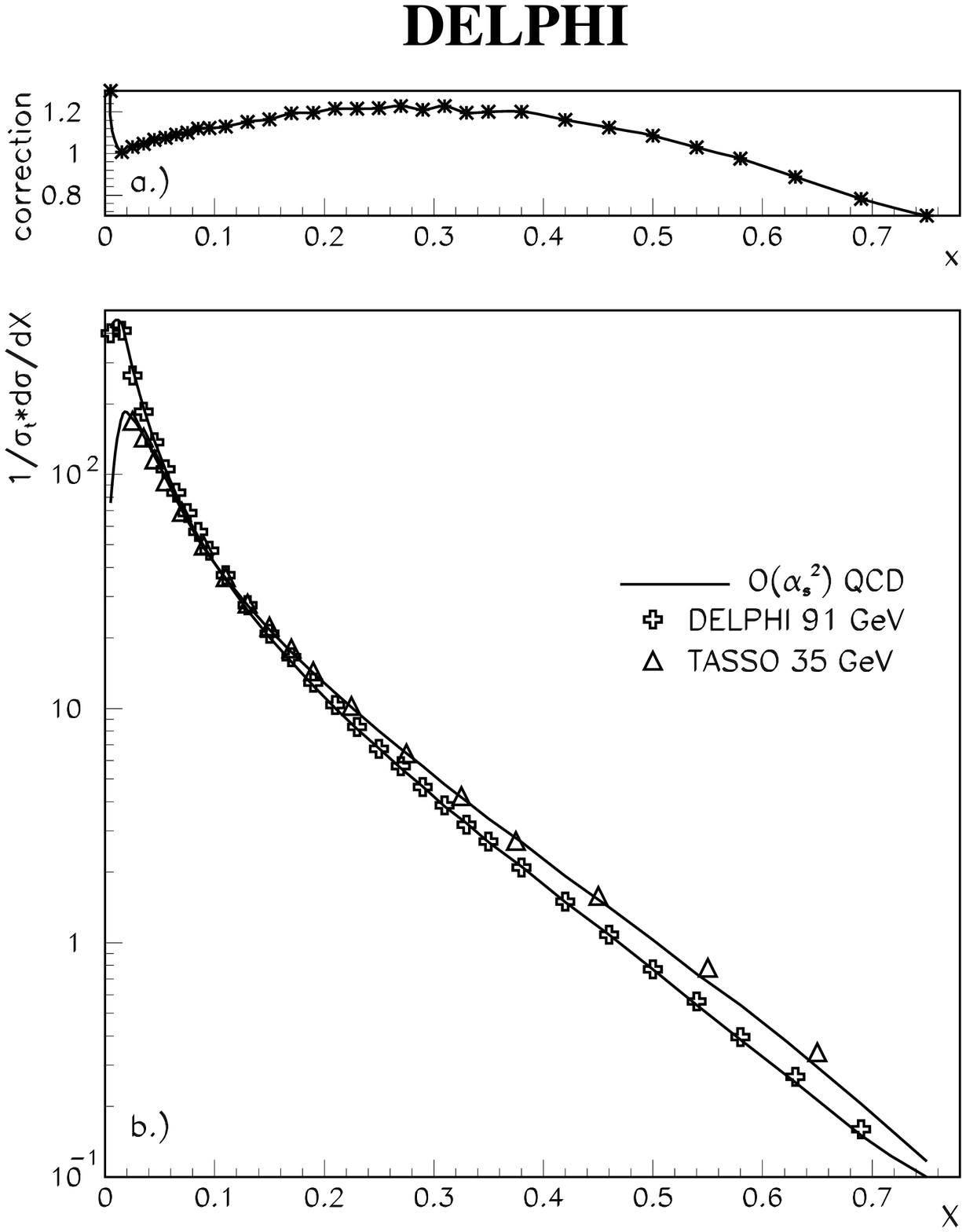}}
\vspace{-0.3cm}
\caption{
a) The correction factor and b) the corrected inclusive momentum
distribution $\frac{1}{\sigma} \frac{d\sigma}{dx}$, where
$x=p_{hadron}/E_{beam}$  from TASSO data at 35 GeV and
DELPHI data at 91.2  GeV. The solid curves are results of the
fits to the second order QCD matrix element.
}
\label{f1}
\end{center}
\end{figure}

The scaled inclusive momentum spectrum
was       obtained by correcting
the $x$ distribution of all charged particles
for initial state radiation,
                          particle decays, detector effects,
and selection cuts.
          In principle $x$          is the
fraction  of the beam energy $ E_{beam} $  carried by the  hadron $h$,
 i.e. $x=         E_h/E_{beam}$,
but  instead of $E_h$ the momentum $p_h$ was used. This was
 experimentally better measured and provided the same
 scaling violation information.
The corrections were obtained from a detailed Monte
Carlo simulation of the detector after generating the  hadronic
\zz\ decays with  the Monte Carlo program from the Lund group\cite{lun}.
Higher order
initial state bremsstrahlung radiation was implemented
by using the DYMU3 program\cite{dym}.
The corrected data were obtained
          by multiplying the data in each bin of a histogram by a
correction factor  defined as
\begin{equation}
c^{(i)} =\frac{N_{gen}^{(i)} }{N_{sim}^{(i)} },
\end{equation}
where
$N^{(i)}_{gen}$ are the contents of the  histogram bin $i$  at the
generator level witout initial state radiation, and $N_{sim}^{(i)}$
after initial state radiation and detector simulation.
       All        primary particles
with a lifetime larger  than $3\cdot 10^{-10}$ s were assumed to be
stable at the generator level, i.e. they were included in
$N^{(i)}_{gen}$, and all those with a shorter mean life (including
  $K_S^0$'s and $\Lambda$'s) were allowed to decay  as part of the
  simulation process. Thus the corrected distributions include the
  contributions of these short-lived particles,  as was the
practice  in   experiments at lower energies.
                        Each histogram was normalized to the
total  number of events.
The corrected      distribution is simply:
\begin{equation}
N^{(i)}_{corr} = c^{(i)} \cdot N^{(i)}_{Data}.
\end{equation}
The corrections varied smoothly and
were less than  35\% for  $x$ below 0.8
 (see Fig. \ref{f1}a).
Only this  range was     used for the comparison with QCD
in the next section. The correction factors
     deviated from one due to acceptance losses and momentum
measurement errors.
For larger $x$ values the  momentum measurement errors
dominated   and,   together with the steeply falling spectrum,
caused a smearing towards higher momenta,
   resulting in a correction factor of    0.4 near $x$=1.
   The  corrected spectrum and thus the fragmentation function
  was obtained from the total number of events $N_{t}$ and the
  corrected event numbers $N^{(i)}_{corr}$ for each $x$ value:
  \begin{equation}
  \bar{D}(x,Q^2) \equiv
            \frac{1}{ N_{t}} \frac{d N_{corr}^{(i)}}{dx}
     \equiv \frac{1}{\sigma_{t}} \frac{d\sigma}{dx} (\epemt h+X),
  \end{equation}
The corrected spectrum is displayed in Fig. \ref{f1}b
and tabulated in Table \ref{t1}   together with the statistical
and systematic uncertainties.

      The systematic uncertainties were determined by varying the
      selection criteria and by using different Monte Carlo
      simulations.
      The largest uncertainty in the correction
      factor is connected with the uncertainty in the
      charged multiplicity:
      the integral of the $x$ distribution is equal to
the total charged multiplicity, since
    each event     has $N_{ch}$ entries, so a systematic
error       in the multiplicity after
detector simulation translates into an uncertainty on the
normalisation of the $x$ distribution.
Varying the cuts, especially changing the  minimum number of charged
particles from 5 to 6 and varying the cut on the sphericity axis
between $30^o$ and $45^o$,  changed the correction factors
by less than 10\% of their deviation from 1, i.e. typically
3\% in the intermediate range, but up to 10\% for $x$ above 0.8
and $x$ below 0.04.

In the LUND Monte Carlo program several generators can be used.
Partons  can be generated either with the ``Parton Shower''
algorithm or by using the exact $O(\as^2)$ QCD matrix element.
The difference in correction factor between these  two options
was less than 1\% in the intermediate $x$ range.
More details can be found in Ref. \cite{fur}.

The  relative systematic uncertainty from the sources mentioned above
 was parametrised in the
following way:
\begin{equation}
\delta^{(i)}_{sys}=max(0.03, 0.1    \mid 1-c^{(i)} \mid)
\end{equation}
This procedure gives a relative error of at least 3\%
for the intermediate $x$ range and increases the error near
the endpoints.
It should be noted that the systematic uncertainties are
correlated between the bins, since a change in
the selection criteria
moved the correction factors for each bin all
in the same direction. These correlations will be taken into
account in the determination of the strong coupling constant.

  A significant scaling
  violation is observed between the
      DELPHI spectrum  at 91 GeV and the data from
            TASSO\cite{tas} at a lower  centre of mass  energy
(see Fig.  \ref{f1}b).
Note that the errors are smaller than the symbols for most of the
data points on this logarithmic plot.
In order to show the  scaling violation and the errors
more clearly, the ratio of the curves in Fig. \ref{f1}b
is shown in Fig. \ref{f2}.
  As can be seen, the scaling violation varies from + 40 \%
 to -30\% for $x$ varying between 0.01 and 0.7.
  The deviation of the data from the horizontal line
  in each $x$ bin is, to first order,
  proportional to the strong coupling constant,
         so each data point yields   an independent measurement
  of \as. All values should be consistent, which is a strong constraint
  and simultaneously a cross check.

\begin{figure}[b]
\vspace*{0.4cm}
\begin{center}
\mbox{\epsfysize=20cm\epsfbox{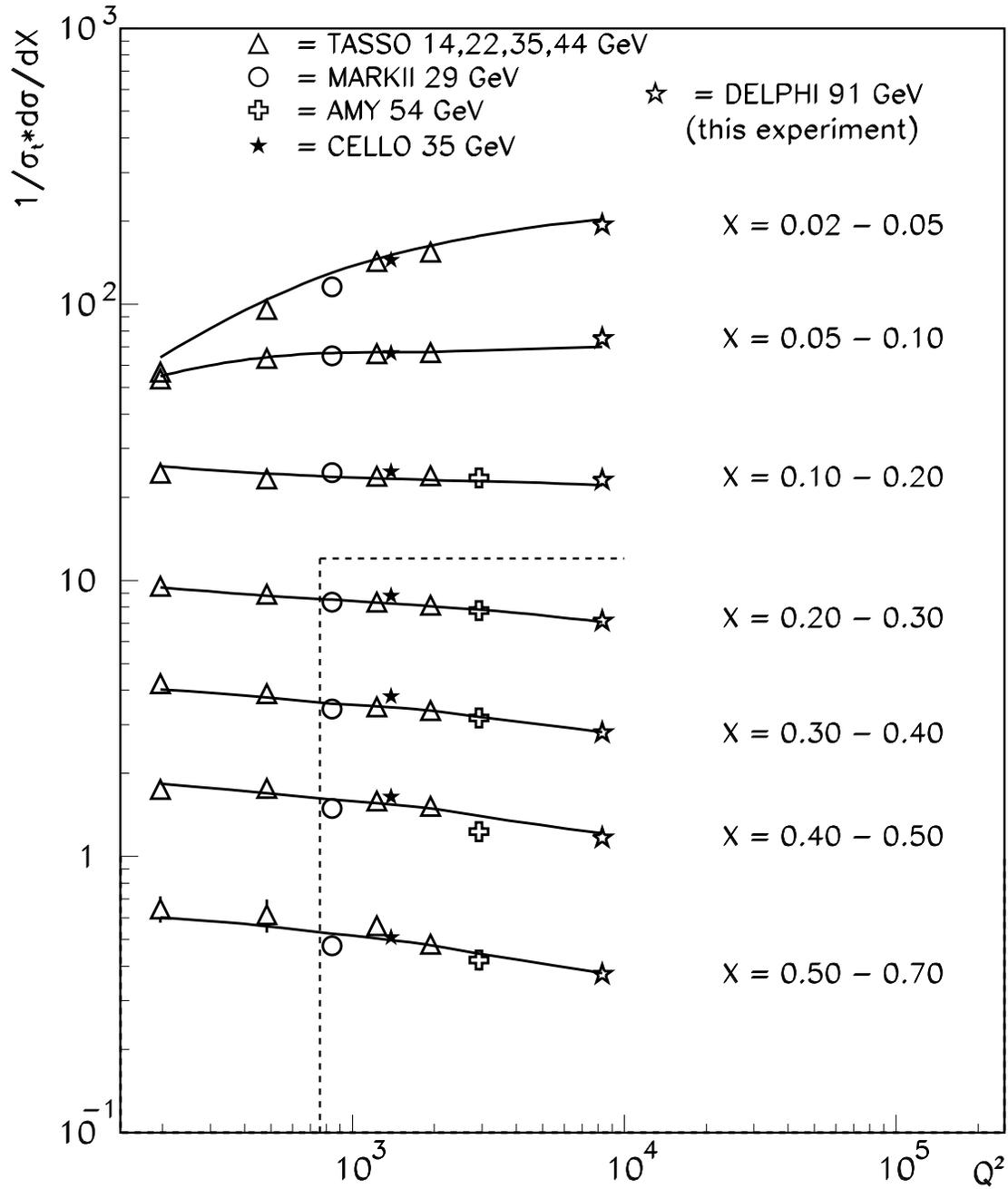}}
\vspace*{-0.3cm}
\caption{
      The $Q^2$ dependence of the inclusive momentum cross section
      in GeV$^2$
for various $x$ bins. For most data points  the errors are
smaller than the symbols.
   The solid curves are results of the fit to
 the data at high $Q^2$ and high $x$,
i.e. the data in the corner (bottom, right)
   indicated by the dashed lines.
}
\label{f3}
\end{center}
\end{figure}

      In Fig. \ref{f3} the DELPHI data are  compared with
other experiments at lower energies\cite{tas,cel,mk2,amy}
 for several        $x$ intervals.
Clearly, at small $x$ values the fragmentation function
increases about 30\%, while at high $x$ values it decreases
by roughly the same amount.
This is exactly what is expected: the higher the energy,
the more phase space becomes  available for gluon radiation. Since the
primary quarks lose more energy due to radiation,
this depopulates the high $x$ region.
The radiated gluons tend to populate the small $x$ region,
increasing the spectrum there.

The curves are the QCD fits for high $Q^2$ and large $x$, as will be
described in the next chapter.
Clearly all $x$ ranges agree well with the QCD fits, even if
they are extrapolated to small $x$
and small $Q^2$. The fact that all regions can be described by a
single value of the QCD scale $\lms$ provides the cross check
mentioned above.

      \section{ Comparison with QCD}

\subsection{Theoretical Framework}
The scaling violations in the fragmentation function,  defined
by Eq. \ref{e1}, are described
by the coupled           integro--differential
evolution equations\cite{alt}, which 
can be written as:
\begin{equation}
Q^2\frac{\partial}{\partial Q^2}
         \left(\begin{array}{c}
\bar{D}_{q}(x,Q^2) \\
\bar{D}_G(x,Q^2)
\end{array}\right)=\frac{\alpha_s (Q^2)}{2\pi}
         \left(\begin{array}{cc}
P_{qq}(z) &  P_{Gq}(z) \\
2\sum  \limits_{i=1}^{5} P_{qG}(z) &  P_{GG}(z)
\end{array}\right)\otimes
\left(\begin{array}{c}
\bar{D}_{q}(x,Q^2) \\
\bar{D}_{G}(x,Q^2)
\end{array}\right) .
\label{ap}
\end{equation}
 The splitting functions $P_{ij}(z)$  in the 2x2 matrix
 are the probabilities of
 finding parton $i$  with momentum fraction  $z$ from its parent parton
 $j$  where $i,j=G$
 refers to a gluon and $i,j=q$ to a quark.
            Note that a gluon can split into a
quark--antiquark pair
of any flavour, hence  the summation and the factor
two in front of $P_{qG}$.   As mentioned before,
 $\bar{D}$  represents the sum over the weighted
contribution of each flavour (see Eq. \ref{e1}), each
having its own fragmentation function $D_i$.

 The  splitting functions can be
 obtained by integrating the exact QCD matrix element.
 In order to obtain the probability of finding a hadron with
 momentum fraction $x$, one has to integrate $P_{ij}(z)$
                    convoluted
 with the probability $D_i(x/z,Q^2)$ that the parton
with energy fraction $z$
 fragments into
 a hadron with momentum fraction $x$.
 The symbol $\otimes$ denotes a convolution integral:
\begin{equation}
P_{ij}(z)\otimes \bar{D}(x,Q^2) \equiv
\int\limits_{x}^{1}\frac{dz}{z}P_{ij}(z)\cdot \bar{D}(\frac{x}{z},Q^2).
\label{e3}
\end{equation}
 Note that  $x/z$ is the fractional hadron energy,
i.e. $x/z=p_{h}/p_{parton}$,   since
       $x=p_h/E_{beam}$ and
     $z=p_{parton}/E_{beam}.$ Obviously, $z$ has to be larger than $x$,
     hence the lower bound in the integral.

    The evolution
equations  describe the $Q^2$   dependence of
             the fragmentation function.
     Their solutions
have not yet been     found in an analytical form.
Numerical solutions, which
account for second
order corrections to the splitting functions  or to   the
anomalous dimensions  have been
developed in Ref. \cite{c22}.
Alternatively, one can integrate the exact second order QCD
matrix element directly, which has some advantages, as will be
discussed later.

In principle the fragmentation of quarks involves an infinite
number of soft and collinear gluons.
Hence, a cut-off on the isolation of the gluons is used
in order to decide whether a gluon should be part of
the quark fragmentation or if it should fragment independently.
In the latter case it contributes to $\bar{D}_G$ instead of
 $\bar{D}_q$.

As a    cut-off,  the minimum invariant mass between
 quarks and gluons was required to be above 9.1 GeV/c$^2$.
This cut presents an arbitrary definition of quarks and gluons,
but it has to be made in any analysis of the scaling violations.
It was  varied  in order to study its effect on the determination
of \lms, as will be discussed in the  section on systematc errors.

\begin{figure}[b]
 \begin{center}
  \leavevmode
  \epsfxsize=16cm
  \epsfysize=12cm
  \epsffile{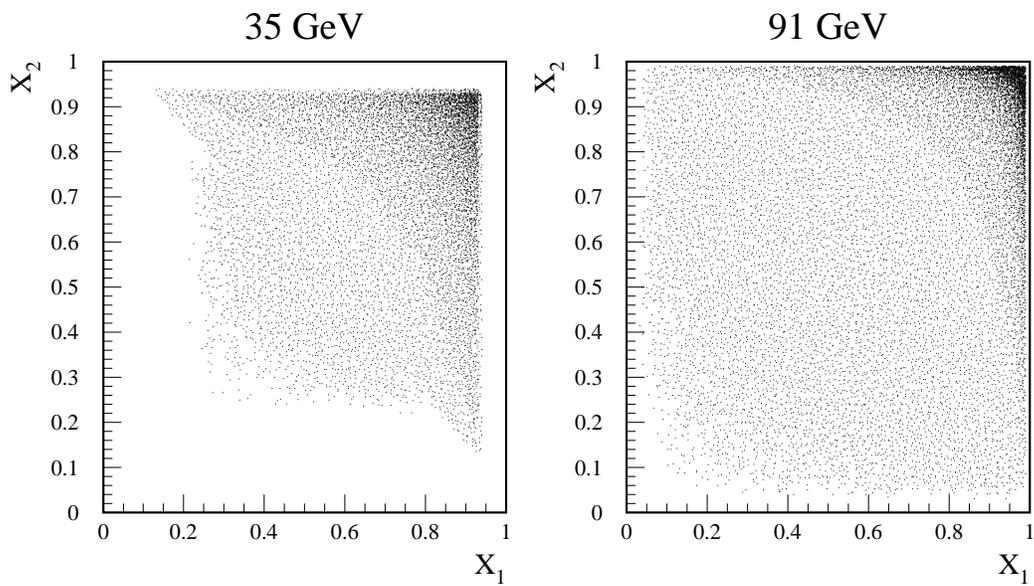}
 \end{center}
\caption{\label{f3a}
The phase space for events with gluon radiation for two
centre of mass energies (35 and 91 GeV) in the $x_1$ versus $x_2$ plane,
where $x_1$ and $x_2$ are the fractional quark energies.
The invariant mass between any pair of partons is required to
be above 9.1 GeV, which causes the empty bands near $x_{1(2)}$=1.
}

\end{figure}

\begin{figure}[b]
\vspace*{0.5cm}
\begin{center}
\mbox{\epsfysize=15cm\epsfbox{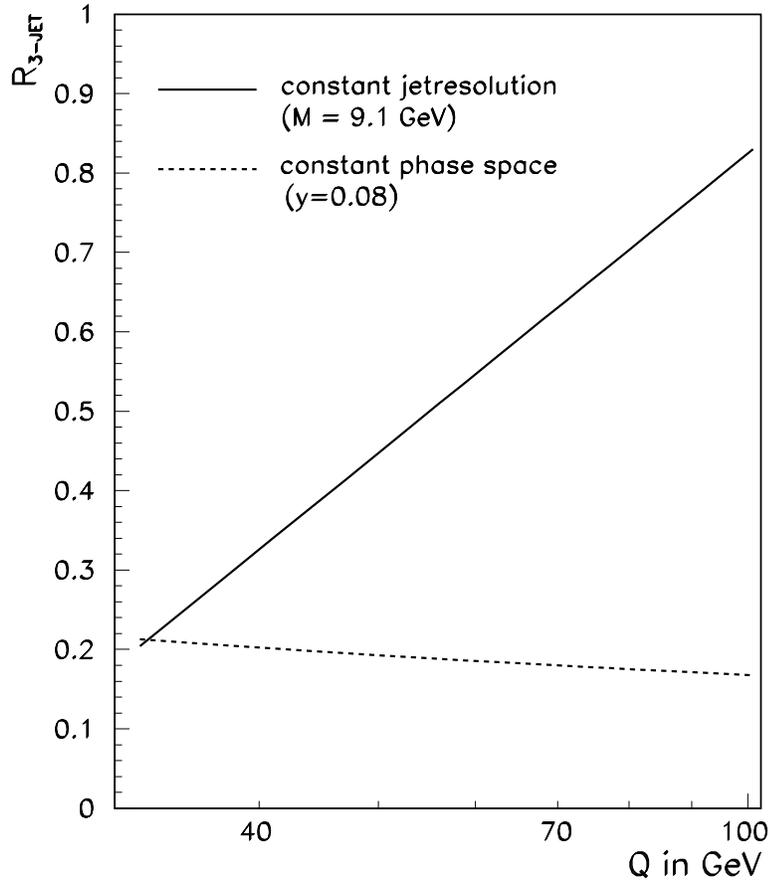}}
\vspace*{-0.5cm}
\caption{
Scaling violation is caused by the $Q^2$ dependence of
the gluon radiation.
This figure shows the two contributions in first order:
the increase in phase space   of the $q\bar{q}G$ 3-jet rate
(solid line) and the running  of the coupling constant, which
causes a decrease of the 3-jet rate, if the fraction of
phase space is kept constant (dashed line).
The phase space was kept constant by requiring at all energies
a minimum scaled invariant mass $y=0.08$ between all partons.
}
\label{f3b}
\end{center}
\end{figure}

Such an invariant mass cut selects a  certain part of phase space,
which varies with energy,
as shown in  Fig. \ref{f3a}.
Here the energy fractions $x_k=E_q/E_{beam}$
of both
quarks in second order QCD were plotted against each other
at   centre of mass energies of 35 and  91 GeV.
The 2-jet events are located at $x_1=x_2=1$ and the 3- and 4-jet events
more towards the centre.
One clearly observes the strong increase in phase space for the
events away from the 2-jet region.
The invariant mass cut
$$y=M_{ij}^2/s=1-x_k,$$
where $M_{i,j}$ are the invariant masses between any pair of partons
and $x_k$ are the fractional quark energies,
eliminates the soft and collinear
gluons in the regions $1-x_k <       M_{ij}^2/s =0.0676 (0.01) $
for the centre of masss energies of 35 (91) GeV.
The difference in phase space
between these energies increases  the $q\bar{q}G$ cross section
with a given `hardness' of the gluon, i.e. with a given invariant mass
cut, by a factor four: the 3-jet rate varies from
   20\% to 80\%, as shown (for a constant value of $\as$ of 0.121)
by the solid line in Fig. \ref{f3b} .

In addition to the $Q^2$ dependence of the
phase space, one has to consider
  the $Q^2$ dependence of $\as$, which has
    the opposite effect: it decreases the $q\bar{q}G$ rate with
increasing energy. This decrease, from the running of the coupling
constant, can be observed if       the phase space for the $q\bar{q}G$
final state is defined as a constant fraction of the total phase space,
 for example by a constant $y$-cut instead of a constant invariant
 mass cut.
In this case the only $Q^2$ dependence
comes from the running of the coupling constant, which
decreases the $q\bar{q}G$ cross section by about 20\% if the
centre of mass energy is increased from 35 to 91 GeV\cite{rev}.
This decrease of the 3-jet rate in a constant fraction of phase space
is shown (for a fixed value of $\lms$ of 270 MeV and renormalisation
scale $Q^2 = s$) as the dashed line in  Fig. \ref{f3b}.
Note that the scaling violation  from the running of \as\
is a small effect compared with the scaling violation from  the change
in phase space, as is apparent from Fig. \ref{f3b}.

  The large phase space dependence can be absorbed in
the fragmentation function,  which then depends on
both $x$ {\it and} $Q^2$.
The redefined
cross section has a well determined perturbative expansion
 in $\as(Q^2)$. This would not be the case  if the large phase space
 corrections, proportional
to $\as \ln Q^2$, were considered to be QCD  corrections.

The energy dependence of \as ~can be expressed in terms of the
energy independent QCD scale $\lms$;  here the upper index indicates
the number of flavours
  $n_f=5$  and the lower index  the
renormalisation scheme (following the convention of Ref.\ \cite{pdb}):
\begin{equation}
\as(\mu^2) = \frac{4\pi}{\beta_0~L}
\left[
  1 - \frac{\beta_1}{\beta_0^2}  \frac{ ln L}{L}
\right]
\label{eq:asm}
\end{equation}
with 
\begin{displaymath}
\begin{array}{ccl}
L      & = &  ln(\mu^2/\lms^2)    \\
\beta_0& = & 11-\frac{2}{3}n_f \\
\beta_1& = & 2(51-\frac{19}{3}n_f)\\
\end{array}
\end{displaymath}
               The energy scale $\mu^2$ of \as\  can be related
to $Q^2=s$ by
$$\mu^2=f s,$$  where $f$ is the renormalisation scale factor.
 The choice of $f$ is free and  QCD predictions  would
not depend on it if all higher orders were known.
In practice,  calculations have been performed only up to a fixed order
and varying $f$ in a wide range indicates the uncertainty due to the
higher orders, as will be discussed in the section on systematic errors.
Note that different choices of $f$ change the value of $\as$.
In order to keep the physical observables constant, the
coefficients of the higher order terms in the
\as\ expansion of the observable have to be changed
correspondingly\cite{rev}.

 The extraction of \as\ from the observed
scaling violations is straightforward. First, the
$x$ dependence  of the fragmentation function,
which cannot be calculated perturbatively,
must be   parametrised from data
at a reference energy. Starting from this parametrisation at the
reference energy,
     the evolution
to  higher energies is predicted by QCD
and compared with the observed fragmentation function
at these energies.
In the following sections the parametrisation of the $x$ dependence
and the $Q^2$ dependence  of the fragmentation functions will
be discussed.

\subsection{Parametrisation  of the Fragmentation Function }
      The fragmentation functions have been studied in great detail
in $\epem$ annihilation. Even such details as the
``string effect'', predicted in QCD by the interference effects of
multiple gluon emission, have been observed\cite{str}
 and can be well described
by   the string fragmentation model\cite{lun}.
 Although any parametrisation of the $x$ dependence at
 a given $Q^2$  would suffice,  we have chosen the  string model
 for the following reasons:
 \begin{itemize}
 \item The  quark and gluon fragmentation functions are described
       by the $\it{same}$ string with the same parameters,
        thus reducing the number of free parameters.
\item In this model        soft  gluons are automatically ``absorbed''
      in the string, i.e. they only produce some
      transverse momentum   to  the string, but do not
      lead to independent jets.
                 In independent fragmentation models
       the fragmentation of soft gluons is problematic
       because of phase space restrictions for hadron production.
\item Quark  mass effects
                   are taken into account in the string model.
\end{itemize}

Hadrons inside a  jet   are characterised by the
   limited transverse   momenta with respect to the
jet axis independent of the jet energy and the longitudinal
momentum spectra. These momentum components
 can be parametrised   by two energy independent functions,
   a longitudinal and a transverse       fragmentation   function.
Italics have been used here  in order to distinguish these
parametrisations   at a reference energy
   from the fragmentation function $\bar{D}(x,Q^2)$.

The transverse momentum spectrum was parametrised by a Gaussian
with a variance of ($500$\ MeV/c)$^2$\cite{deb}.
The longitudinal momentum spectra of light and heavy quarks
are parametrised differently since the latter have much harder
spectra because of their larger mass.
The  Lund symmetric $fragmentation$ function\cite{lun0} was used
for the light quarks:
\begin{equation}
f(y)=\frac{(1-y)}{y}^a \exp\left(
- -\frac{b\cdot m_{\perp}^2}{y}\right)  ,
\end{equation}
where $m_{\perp}=\sqrt{m^2+p_{\perp}^2}$ is the transverse mass
of the hadron,    $y=(E+p_l)_h /(E+p_l)_p$
determines the
    fraction of the primordial parton       energy
       taken by the hadron $h$, with $p$, indicating the parton
with energy $E$ and longitudinal momentum $p_l$;
           $a$ and $b$ are two free parameters.
                                         The longitudinal
spectrum depends mainly on
 $a-b$ which scales like $N$,
where $N$ is the total multiplicity,
so effectively there is only one free parameter.
For the heavy quarks ($b$ and $c$)              the Peterson
$fragmentation$    function\cite{pet}    was used:
\begin{equation}
f(y)=\frac{1}{y}\left[1-\frac{1}{y}-\frac{\epsilon_i}{1-y}  \right]^{-2}
\end{equation}
Here the free parameter, $\epsilon_i$, is expected to vary
as $1/m_q^2$, so               $\epsilon_c/\epsilon_b=9.4$ was used.
Hence there are only two free parameters to tune the
momentum spectrum ($a$ and $\epsilon_b$).
      The parameters determining  the fraction of
strange quarks picked up from the vacuum,    the ratio of vector
to pseudoscalar mesons, the fraction of baryons, as well as
the decay parameters were all left at their default values,  since
a good description of the $x$
dependence
was possible with these.


   \subsection{    Determination of the Strong Coupling Constant}
The  $Q^2$ dependence of the fragmentation function        can be
 derived either from  the evolution equations or from a direct
numerical integration of the exact QCD matrix element.
Since the splitting functions have been derived from the integration
of the matrix element, both methods are, in principle, equivalent.
However, higher order differences might occur. As will be discussed
in the section on systematic errors, these differences are small.
Therefore,
   the $Q^2$ dependence has been determined from the integration of the
exact
second order QCD matrix element, using the formulae from
reference \cite{ert},
which have been  implemented in the Lund string model\cite{lun}.
This method has the advantage that the convolution of the
splitting and fragmentation functions is done in a consistent manner,
i.e. the cuts to separate the nonperturbative          region from
the perturbative one are the same for the splitting and fragmentation
functions.  These cuts will be discussed in more detail in the
section on systematic errors.
Furthermore, the weights $W_i(Q^2)$ in Eq.\ \ref{e1}
from the electroweak theory have been incorporated in this model.

The strong coupling constant was  extracted in the following way.
A simultaneous fit of the QCD scale $\lms$ and the
fragmentation parameters $a$ and  $\epsilon_b$
was made by minimising:
\begin{equation}
\chi^2=\sum\limits_{j}\left[
         \sum\limits_{i}\left(
         \frac{(f_j     D^{(i)}-T^{(i)})    ^2}{(f_j
\sigma_{exp}^{(i)})^2}\right)
 +\frac{(1-f_j)^2}{\sigma_n^2}\right],
 \label{chi}
\end{equation}
where $f_j$ is the normalisation factor for experiment $j$ with
    data                   $D^{(i)}$ in a given $x$ bin with an experimental
    error $\sigma^{(i)}_{exp}$ for that bin
     and an overall normalisation error
$\sigma_n$.       The fit function $T^{(i)}$ was the  $x$
parametrisation from the   string model convoluted with
the $Q^2$ dependence from the integration of the exact QCD
matrix element 
 and the $Q^2$ dependence of \as.
In order to prevent a bias from b-quark
threshold corrections, only data
                     at or       above $E_{cm}$=29 GeV were      used   .
Furthermore, data at high and low  $x$ values have not been used, since
the experimental correction  factors are large in these regions.
Fitting the data from Delphi simultaneously with all other available
data\cite{tas,cel,mk2,amy,ale}
 in the range $0.18<x<0.8$ and $29^2<Q^2<91.2^2$ GeV$^2$
yielded the results given in Table \ref{t0}.

The fit was repeated for two values of the renormalisation scale.
For $f=\mu^2/s=0.01(1.0)$ the result was:
$\lms=193^{+20}_{-11}$~ $(269^{+17}_{-14})$  MeV.
%
The  fit results   were obtained for a value of $b=0.283$ in  the Lund
symmetric $fragmentation$ function\footnote{Although $a$ and $b$
are        strongly correlated, one   could not leave $b$ at an arbitrary
value and just fit $a$ or vice-versa. A good parametrisation was
obtained if $b$ was chosen in a range around 0.3.}.
A good agreement was observed for all $x$ values  with
the $\it{same}$ fragmentation parameters at both 35 and 91 GeV, so the
difference between the energies depended       on $\lms$ only.
                       The results for $f=0.01$
are shown as the solid  lines in Fig. \ref{f1}b; the $\chi^2$ of the
fit for $f=1.0$  was equally good.

\subsection{Systematic Uncertainties}
\begin{figure}[b]
\vspace*{1.0cm}
\begin{center}
\mbox{\epsfysize=18cm\epsfbox{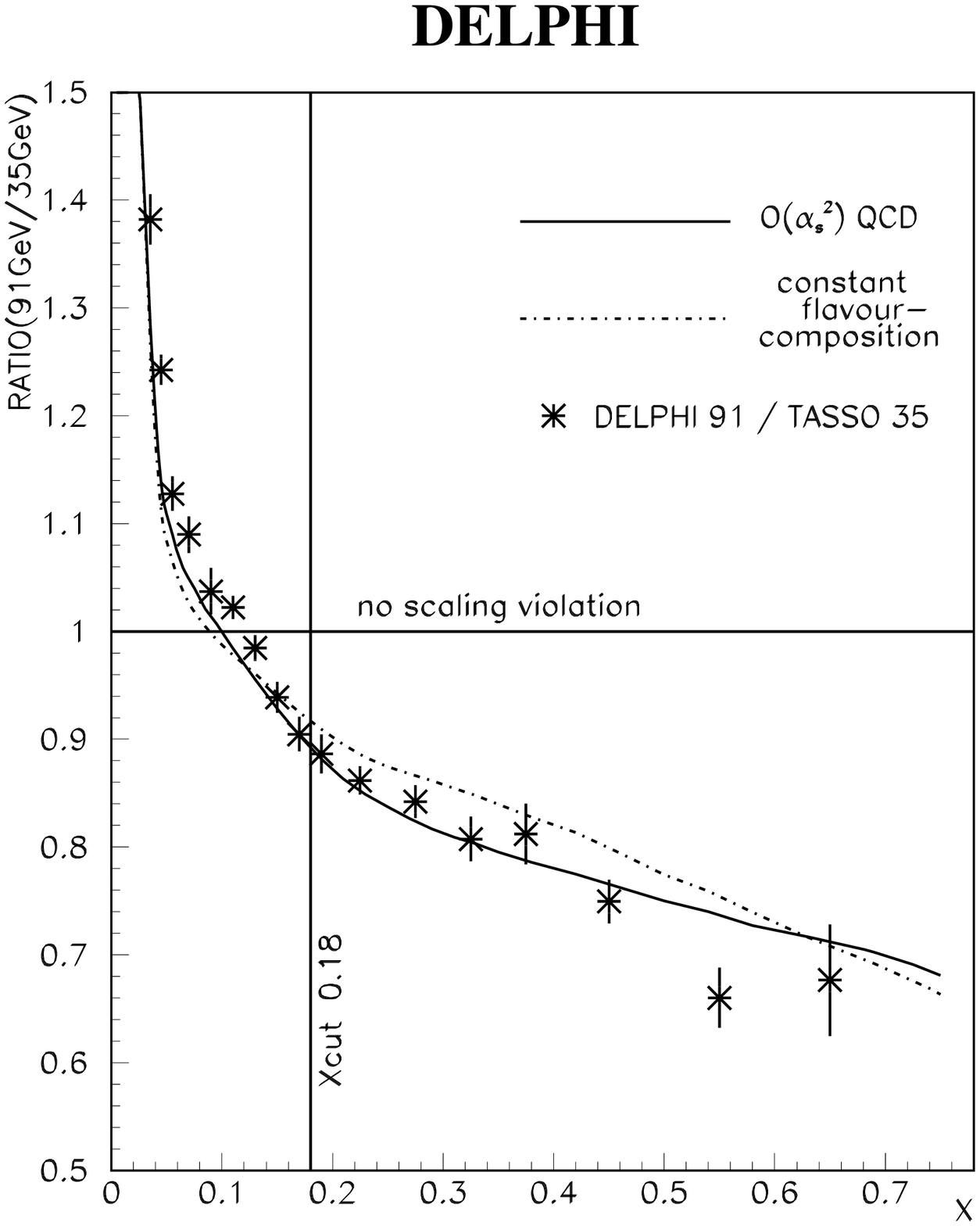}}
\vspace*{-0.2cm}
\caption{
The ratio  of the   curves in Fig. 1b.
 The dashed-dotted
  line assumes that the flavour composition  at 91.2 GeV  is the same
as the one at              35 GeV.
As is apparent from the small difference between the solid and
dashed-dotted lines, the increase in heavy quark production at the
\zz\ resonance does not influence the scaling violation strongly.
The reason is simply that although the heavy quark fragmentation is
 harder, the momentum spectra {\it after} the decays
look similar to the ones from
the light quarks and the difference does not show
the characteristic energy dependence from the scaling violation.
}
\label{f2}
\end{center}
\end{figure}

\begin{figure}[b]
\vspace*{0.5cm}
\begin{center}
\mbox{\epsfysize=20cm\epsfbox{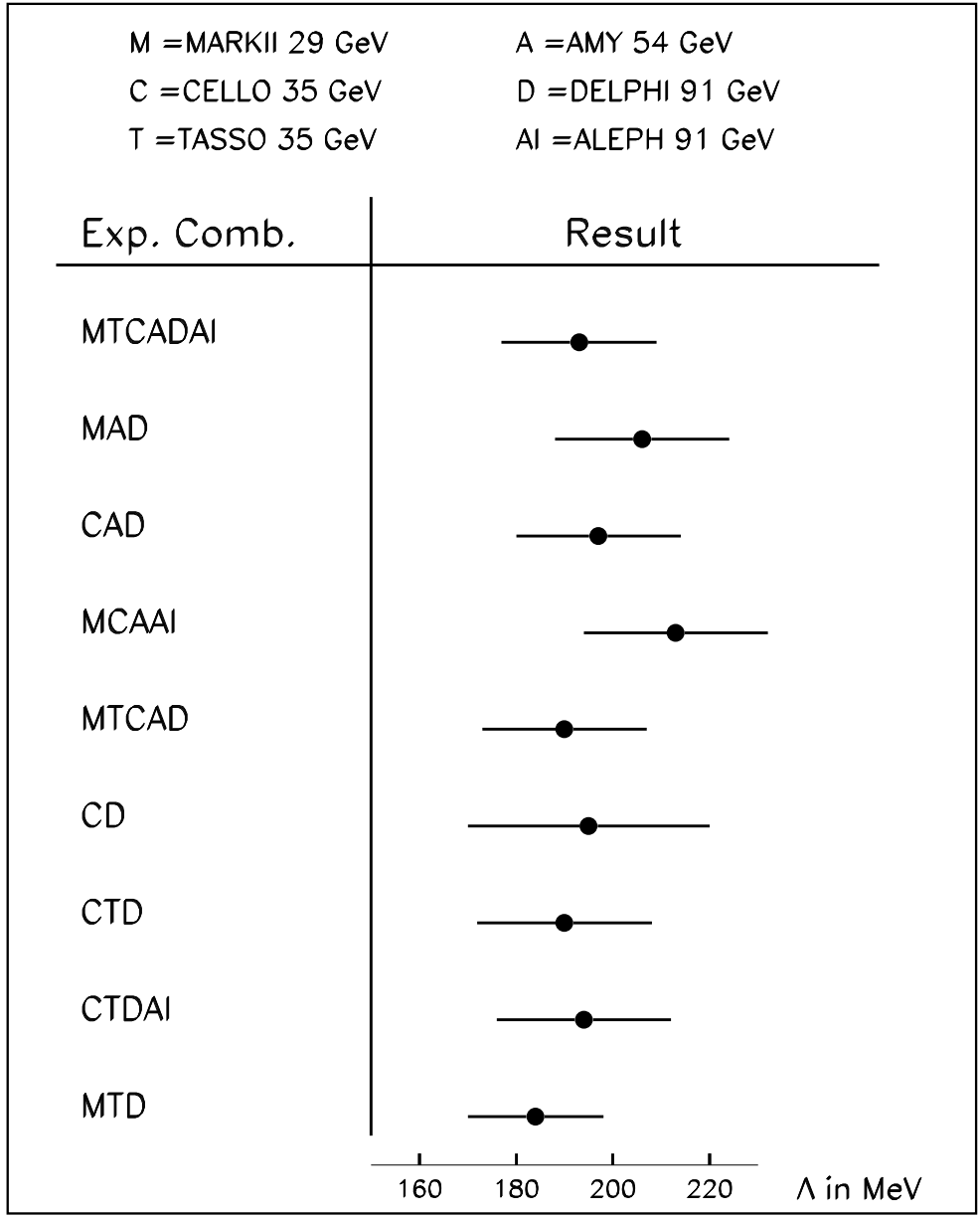}}
\vspace*{-0.5cm}
\caption{
      The \lms\ values obtained from fitting various combinations
of experiments.   The  combinations were choosen such that the
statistical errors were similar. They
are indicated on the
left using the abbreviations given at the top.
}
\label{f4}
\end{center}
\end{figure}

\begin{figure}[b]
\vspace*{0.4cm}
\begin{center}
\mbox{
\hspace*{0.1cm}
\epsfysize=20cm\epsfbox{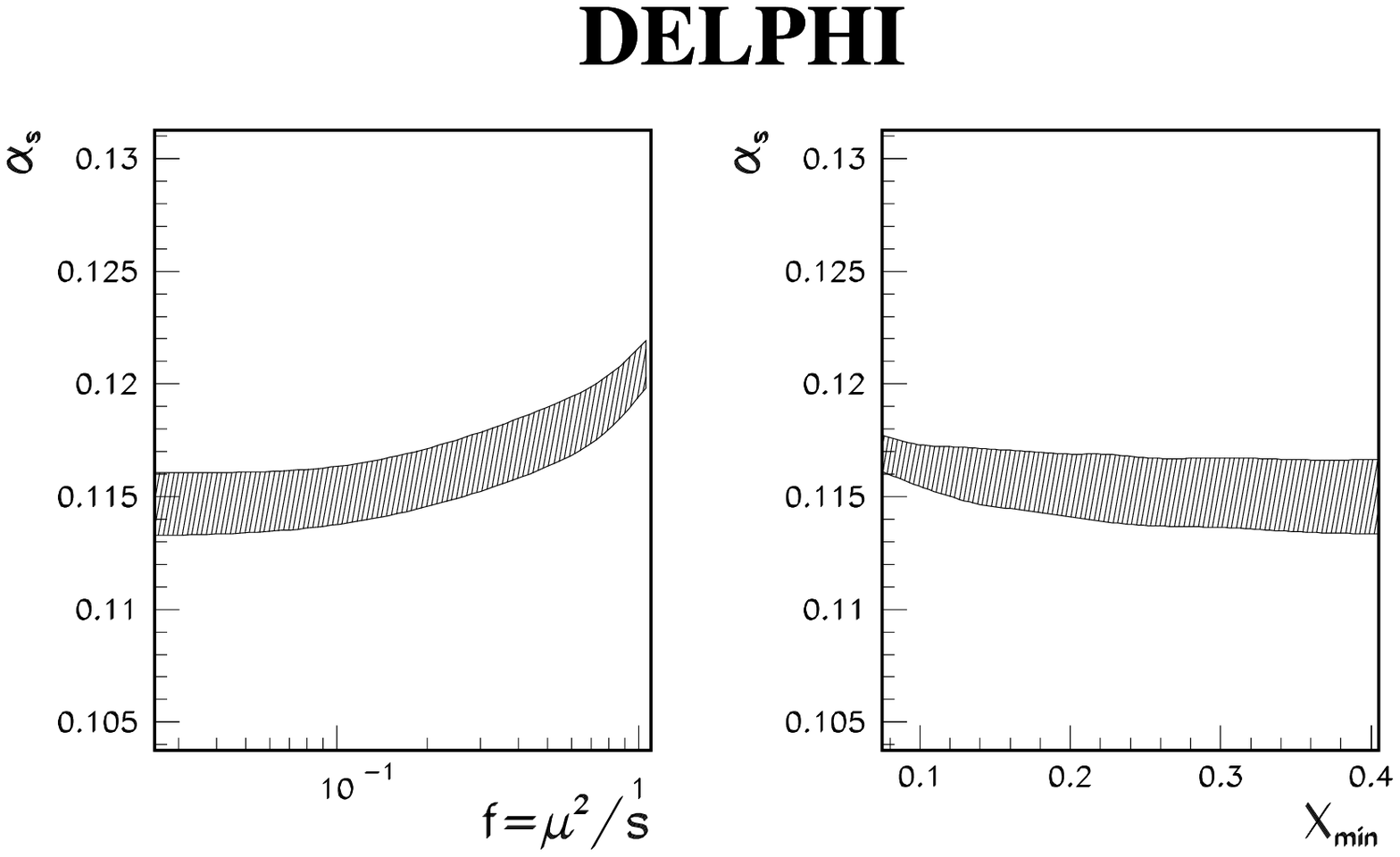}}
\vspace*{-3.5cm}
\caption{
The renormalisation  scale dependence of \as\ (left) and the
$X_{min}$ dependence (right).
$X_{min}$ is defined as the
the minimum value of $x$ used in the fit.
}
\label{f5}
\end{center}
\end{figure}

 The  results in the previous section include  both
    systematic and statistical uncertainties, as well as
 the uncertainties from the correlation between the fragmentation parameters
               and $\lms$.      In addition there are theoretical
 uncertainties from the unknown higher order corrections, which
 are usually estimated by varying the renormalisation scale.
To get the complete error estimate, the
following
have been investigated:
\begin{itemize}
\item
  \underline{Experimental uncertainties.} \\
In the definition of $\chi^2$, Eq. \ref{chi},
$\sigma^{(i)}_{exp}$ represents
the total error for that data point, obtained by  adding in
quadrature the statistical and point-to-point systematic uncertainty,
but excluding the overall normalisation  error, $\sigma_n$.
However, the separation between point-to-point systematic
uncertainty and normalisation uncertainty is not straight forward   and
usually not given in the literature.
Furthermore, the published systematic uncertainties are not always
comparable in the possible sources which have been included.
Therefore  the systematic uncertainties were varied considerably
in order to  check their influence on the fitted value of \lms.
The following procedure was adopted:
the total error, $\sigma_{tot}$, was split into a point-to-point error
$\sigma^{(i)}_{exp}$ and a normalisation error $\sigma_n$:
\begin{equation}
  \sigma_{tot}^2={\sigma^{(i)}_{exp}}^2 + \sigma_n^2;
\label{sig}
\end{equation}
$\sigma_n  $ was varied from 1\% to 3\% and subtracted from the
total error          quoted by the experiments (using Eq. \ref{sig}).
If the remaining point-to-point error fell below a certain
minimum value, it was adjusted to this minimum value in
order to ensure that the point-to-point error squared would not become
negative  or too small for experiments in which
all possible systematic effects had not been included
in the error estimate.
This minimum value $\sigma_{exp}^{min}$
was varied between    1\% and 3\%.
Of course, the $\chi^2$ of the fit was changed if the errors were
changed, but fortunately $\lms$ varied by only $\pm 3$ MeV
if     $\sigma_n$ and $\sigma_{exp}^{min}$
             were changed in the ranges given above.
These small changes in \lms\ indicate
that the scaling violations are determined  only by the shape
of the distributions, not the absolute normalisation.
The $\chi^2$ values have been summarized
 in the Tables  \ref{t1}-\ref{t6} using $\sigma_n$=2\% and
 ${\sigma_{exp}^{min}}  $ = 3\%.
 The total $\chi^2$ is 73 for 71 data points in the fit region
 using a renormalisation scale factor of 0.01.
 For comparison,      the $\chi^2$ values outside the fit
 range are shown too.
The correlations between the parameters depended on the assumed
errors, but were never larger than 40\% for any pair of parameters.
The fit normalisation factors were
consistent with one for all experiments, as indicated in the
captions of Tables \ref{t1} and \ref{t2}-\ref{t6}.
\item
 \underline{Differences between experiments.}\\
In contrast to the deep   inelastic lepton scattering experiments,
which measure the $Q^2$ dependence in a single experiment, we had to
combine      data from  different accelerators, so
            systematic effects
from differences   between  experiments had to be considered.
They were checked
 by comparing the results of different combinations of experiments.
                 The maximum difference    in $\lms$ from the
various combinations of the 6 experiments was less than
   30 MeV,  which is not much larger than the statistical uncertainty
(see Fig. \ref{f4}).
              The
systematic uncertainty from this source
was conservatively estimated to be half
the maximum difference, i.e. 15 MeV, thus assuming that the
whole difference was systematic and not due to statistical fluctuations.
The reason for this surprisingly small spread is simple:
all experiments used large $ 4\pi $ solenoidal detectors in which
the momentum spectrum, especially the shape, was clean and easily measured.
As mentioned before, it is the change in the $\it{shape}$ of the
$x$ spectra which determines   the scaling violation,
 not the absolute normalisation.
\item
 \underline{x-dependence.}\\
 For low   $x$ values the contributions from multiple soft gluon
emission start to dominate. In this region
 the $\chi^2$ of the string model
 parametrisation  becomes  somewhat worse  (see Tables
 \ref{t1} and \ref{t2}-\ref{t6}).
To estimate the uncertainty from the small $x$ range, we fit
     between $x_{min}$ and $x_{max}$ and varied $x_{min}$
between 0.08 and 0.4.
Since the experimental correction factor for high momentum particles
becomes large for $x>0.8$, $x_{max}$ was kept at 0.8.
For  $x_{min}$ =0.08, $\lms$ for $f=0.01$ increased
from 190 to 210 MeV, but
for the fit range considered
($x_{min}> 0.18)$,
 no variation in \as\ was seen (see Fig. \ref{f5}).
 Nevertheless,
the  uncertainty for  the selected
 $x$ range was conservatively
estimated to be 10 MeV, which is
half the difference between the
values obtained for $x_{min}=0.08 $ and $x_{min}$ =0.18.
As mentioned in the introduction, each $x$-value provides an independent
determination of \as.
The fact that \as~ is practically independent of the selected $x$-range
indicates that all  $x$-values are consistent.
\item
 \underline{Heavy Quark Fractions.}\\
                     The fragmentation effects largely cancel
in the difference between the spectra at different energies.
However, since the primary quark composition changes with energy,
the influence of
                   the difference in fragmentation between
light and heavy quarks should be considered.
Although the primary mesons from heavy quarks have the hardest
momentum spectra, the spectra after decays are not much different
from the ones for the light quarks
and actually somewhat softer.
Furthermore, it was not possible
to mimic the characteristic change in shape          from
the QCD scaling violations by the difference in quark
compositions, as shown by the dashed-dotted line in Fig. \ref{f2}.

      Fitting the $x$ spectra at 35 and 91 GeV
simultaneously was a good way to determine the
fragmentation of   both light and heavy quarks, since the
different quark compositions at the different energies, combined with
the somewhat softer $x$ spectrum of the heavy quarks,
yielded only a moderate correlation
between  the fragmentation
parameters $a$ and $\epsilon_b$ (see Table \ref{t0}).
The fitted value of the
latter parameter  gave an average $x$ of the B-hadrons of
$0.69\pm 0.01$ at the LEP energy,  which
is in good agreement with the value obtained from
lepton spectra in semi-leptonic  B decays\cite{rou,chr}.
Note that the  determination of $\epsilon_b$ from the inclusive
hadron spectra included all decays
and was therefore independent of the value determined
from the lepton spectra.

 As an additional check that the different  heavy quark fractions at
35 and 91 GeV do not mask the scaling violation from QCD,
the scaling violation was calculated with a constant fraction of
heavy quarks ($\approx 11\%$
 for b-quarks and $\approx 44 \%$ for c-quarks,
which are the values at \sqrts=35 GeV).
 The amount of scaling violation is not changed significantly,
 as shown in Fig. \ref{f2};   the small difference
                             was taken into account in the fit
and the residual uncertainty in $\lms$ was estimated to be 10 MeV.

\item \underline{Independent versus String Fragmentation.}  \\
In the fit, the string fragmentation model was used to parametrise the
$x$ dependence. As an alternative, the independent fragmentation
option in the Monte Carlo program from the Lund group has been
used\footnote {The following parameters were used: f=0.01,b=0.283,
$p_{t}$=500 MeV/c, stopping point of fragmentation 3.1 GeV/c.
The gluon was splitted into light quark pairs according to the
Altarelli-Parisi splitting function and each quark fragmented then
independently. The fitted parameter values were: $a=0.88\pm{0.05}$,
$\epsilon_{b}=0.007\pm{0.001}$,
$\Lambda_{\overline{MS}}=211^{+36}_{-24}$ MeV.}.
In this case all quarks and gluons fragment independently.
The whole analysis, including the parametrisation of the
$x$ dependence, was repeated with this model.
The fit quality was similar  and
the value of $\lms$ was not changed outside the experimental errors,
again indicating that fragmentation uncertainties largely cancel
in the difference between the spectra at different energies.
Half the difference between the different fragmentation models
(9 MeV) was taken conservatively as the error for
fragmentation.
\item
 \underline{The renormalisation scale uncertainty.}\\
 As mentioned before, the renormalisation scale is a free parameter;
\lms\ would be independent of the choice of this scale if all
higher order corrections were known. However,
in a fixed order calculation a
  lower scale implies a larger value of \as. For the 3-jet cross
section the  change in the Born cross section
can be compensated  by a different coefficient in the higher order
correction. However,     the 4-jet cross section is only known
up to the Born term in second order QCD, so a lower scale
for the argument of \as\ implies a higher 4-jet rate.
The $\chi^2$ of the fit did not change
significantly by changing the scale, but \lms\ varied   from
190 to 270 MeV if the scale was changed from $E_{cm}/10$ to $E_{cm}$,
              which corresponded to a change in \as\
from 0.115 to 0.121 (see Fig. \ref{f5}).
Thus this error, originating from
the unknown higher order corrections, has been
found to be dominant, as in all other \as\ determinations\cite{rev}.
The scale dependence was still relatively small, since we studied the
difference between the spectra at different energies, so higher
order contributions and fragmentation effects partially cancel.
Similar observations hold for other ``difference'' variables,
like the Asymmetry in the Energy-Energy Correlations (AEEC) or
the difference in jet masses\cite{rev}.

                     An independent estimate of the
higher order contributions  can be obtained from
the equations (\ref{ap}):
in these equations the higher order terms are taken into account
by exponentiating the leading logarithms proportional to
                              $(\as/2\pi)^n \ln^n Q^2$, which appear
 as leading terms in a calculation to order $n$.
The difference of these terms between $Q=Q_{min}$ and $Q=Q_{max} $ is
proportional to
       $(\as/\pi)^n \ln^n (Q_{max}/Q_{min})$.
Since in our case the difference in $Q_{max}$ and $Q_{min}$ is
only a factor three, the exponentiated form of the leading
logarithms will be close to its second order expansion,
so the higher  order contributions are expected to be small.
This can be checked explicitly  by integrating these equations
in $n$ steps.
Since at    each step a gluon can be emitted,
 this corresponds  to summing up all
higher  order terms   proportional to
                              $(\as/\pi)^n \ln^n Q^2$.
The change in scaling violation between 30 and 90 GeV   was found to be
less than 5\% if $n$ was varied between 2 and 20, so this change
is similar to   the uncertainty from the scale dependence.
Since this exercise was done only in first order, using the program
from reference \cite{web},
    the larger   range from the scale dependence
was used as an estimate of the error
from the unknown higher order corrections.
\item
 \underline{Cut-off dependence.}   \\
 As mentioned before,
 the fragmentation of quarks involves a large
number of soft and collinear gluons.
Hence, a cut-off on the isolation of the gluons was used
in order to decide whether a gluon should be part of
the quark fragmentation or if it should fragment independently.
 Below this cut the quarks and gluons were considered to
 fragment into a single jet, i.e. they were considered to belong to the
 non-perturbative regime in the model and were recombined beforehand.

As a    cut-off,  the minimum invariant mass between
 quarks and gluons was required to be above 9.1 GeV/c$^2$.
                         The  scaling violations were not very
sensitive to this cut, since they just required a different
parametrisation of the nonperturbative part for a different cut.
What mattered  was  a good parametrisation
of the $x$ dependence.
           The cut   could not be decreased, since
           with this cut practically all
      phase space was already used at the highest energy, as shown
           in Fig. \ref{f3a}.
  Decreasing the cut  further would cause the 4-jet  cross section
           to become so large and positive, that
           the 3-jet cross section would become negative in some
           regions of phase space due to the large and negative virtual
             corrections in the second order QCD matrix element
             in that case.
Increasing the minimum invariant mass squared
                     by a factor two
           resulted in an increase of $\lms$ of 60 MeV.
Therefore  an error of $\pm 30 $ MeV was attributed, although  part of
this was presumably  already
       absorbed         in the scale error:
       increasing the cut-off or increasing the renormalisation scale
       increased $\lms$ in both cases, as expected for a decrease
       of the higher order contributions from multiple gluon radiation
       in both cases.
       This can be seen in Fig. \ref{f6}, which shows the phase space
       of 4-jet events at an energy of 91 GeV for the
       default parameters ($f$=0.01 and $y_{min}$=0.01) at the top
       and the changes if $f$ is increased to 1 and $y_{min}$
       to 0.02. For these cases the total 4-jet rate decreases
       from 28\% at the default value to $12$ and 9\%, respectively.
       Note that the change  from the change in the $y_{min}$ cut
       is not only at the border, as one might naively expect.
       This can be understood, if one realizes that
             the cut converts many 4-jet
       events into  3-jet events, which  can be located anywhere
       in the allowed phase space. Consequently,  the variations of
           $y_{min}$     and     $f$ change the higher order corrections
           in a vary similar manner and by counting them as
           independent contributions to the total systematic error
           is a conservative attitude.
\end{itemize}
The total        errors   were
 obtained by adding in quadrature the errors from the fit
                           $(^{+20}_{-11}$\ MeV),
 from the $x$-dependence (10 MeV), from heavy quark fractions (10 MeV),
 from fragmentation (9 MeV),
    from the comparison between experiments (15 MeV),
           from the gluon cut-off dependence (30 MeV) and   from the
scale dependence (40 MeV).
A summary of the systematic errors is given in Table \ref{t7}.
\section{Summary}
A precise determination of the fragmentation function  in the
decay of the $\zz$ boson has been presented.
A comparison with the fragmentation functions at lower energies
shows a strong scaling violation, which leads in second order
to a QCD scale \lms\ between 190 and 270 MeV (see Table \ref{t0}).
Taking the average as the central value and
using the total uncertainties as given in Table \ref{t7}
           resulted  in
$\lms  = 230\pm 60 $ MeV,  which corresponds to
 $$\as(\mz)  = 0.118\pm         0.005 .$$

 These results in the time-like region are in good agreement
with the results on scaling violation from deep inelastic
lepton nucleon scattering (space-like region;
 $\as=0.112\pm 0.005$ \cite{vir,mar,qui})
and with other $\as$ determinations
at LEP from jet rates and shape variables
($\as=0.120\pm 0.007$ \cite{lep,rev}).

\begin{figure}[b]
\vspace*{0.1cm}
\begin{center}
\mbox{\epsfysize=20cm\epsfbox{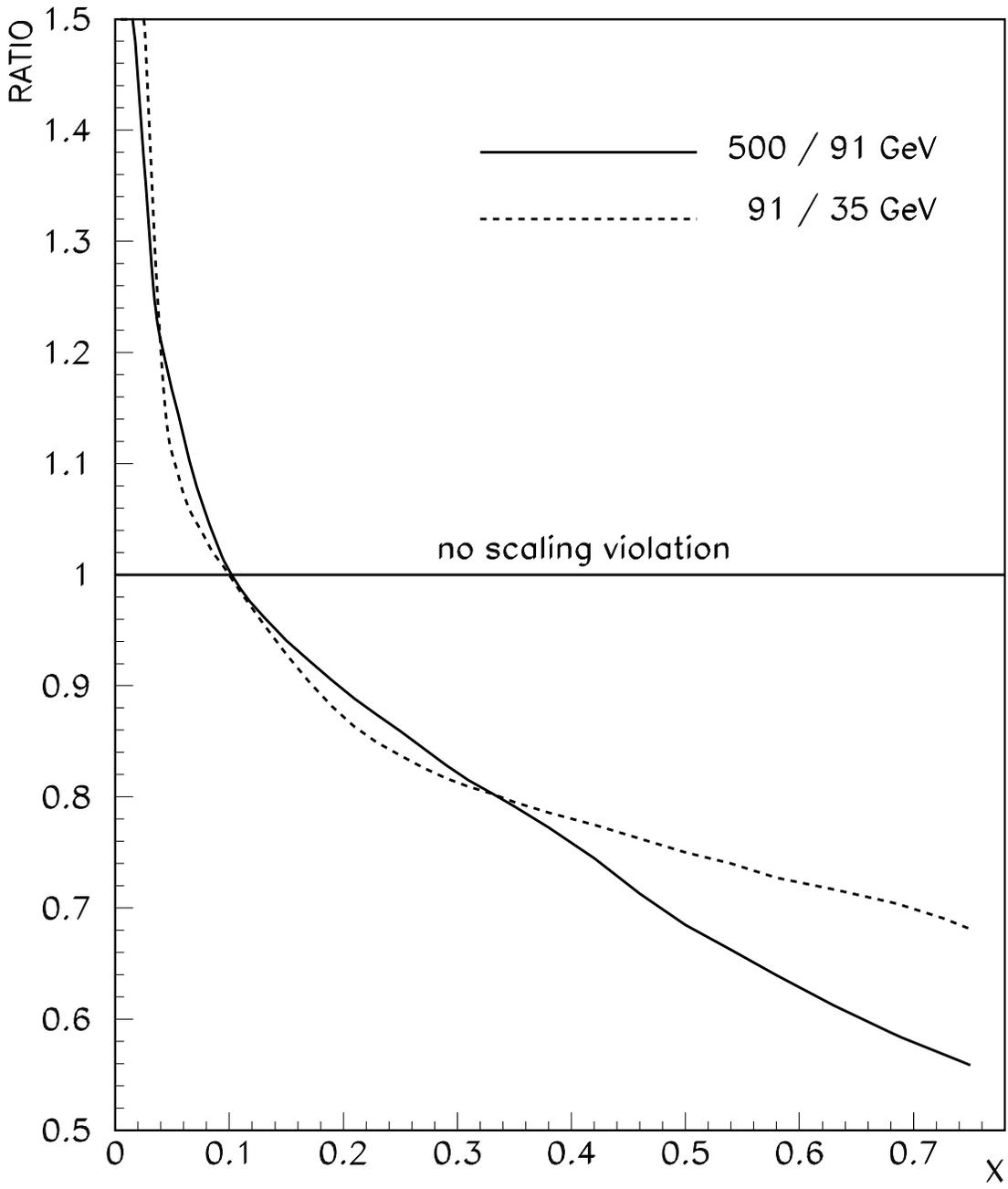}}

\vspace*{-0.4cm}
\caption{The ratio of the momentum spectra at energies of 500 and 91
GeV compared with the same ratio  at energies of 91 and 35 GeV.}
\label{f7}

\end{center}
\end{figure}

It should be noted that future $\epem$ linear colliders
can operate over a large energy range, which allows a study of  the
scaling violations with a $single$ detector. For example,
a 500 GeV linear collider can measure a large scaling violation
by running for a short time ath the 91 GeV $\zz$ peak.
{}From Fig. \ref{f7} it can be seen that the expected scaling violation
will be even larger than the one observed between 35 and 91 GeV,
thus allowing a precise determination of $\lms$ at high $Q^2$
within a single experiment.

\newpage
\section*{Acknowledgments}
 We are greatly indebted to our technical staff, collaborators and
 funding agencies for their support in building the DELPHI detector and
 to the members of the
 LEP Division for the superb
 performance of the LEP machine.

Furthermore, we would like to thank G. Altarelli,
T. Sj\"ostrand,
and B. Webber   for useful discussions and B. Webber  for providing
us with a program for the numerical integration of the Altarelli-Parisi
equations in first order.

\newpage

\begin{table}
\begin{center}
\begin{tabular}{|c|c|c|c|c|c|}
\hline
x-bin &$\chi^2$ &Data &$\sigma_{stat}$&$\sigma_{sys}$&
QCD+SF \\           \hline
0.00 -- 0.01 &   1.38 & 400.8  &  0.8  & 12.1  & 412.9  \\ \hline
0.01 -- 0.02 &   5.60 & 409.3  &  0.7  & 12.3  & 436.2  \\ \hline
0.02 -- 0.03 &   5.27 & 264.6  &  0.6  &  7.9  & 281.5  \\ \hline
0.03 -- 0.04 &   2.01 & 185.1  &  0.5  &  5.6  & 192.0  \\ \hline
0.04 -- 0.05 &   0.39 & 137.4  &  0.4  &  4.1  & 139.3  \\ \hline
0.05 -- 0.06 &   0.01 & 105.3  &  0.4  &  3.2  & 104.5  \\ \hline
0.06 -- 0.07 &   0.21 &  83.6  &  0.3  &  2.5  &  82.0  \\ \hline
0.07 -- 0.08 &   1.00 &  68.4  &  0.3  &  2.1  &  66.0  \\ \hline
0.08 -- 0.09 &   3.46 &  56.9  &  0.3  &  1.7  &  53.4  \\ \hline
0.09 -- 0.10 &   2.12 &  47.2  &  0.2  &  1.4  &  44.9  \\ \hline
0.10 -- 0.12 &   2.68 &  37.1  &  0.2  &  1.1  &  35.1  \\ \hline
0.12 -- 0.14 &   1.25 &  27.6  &  0.1  &  0.8  &  26.6  \\ \hline
0.14 -- 0.16 &   0.53 &  20.9  &  0.1  &  0.6  &  20.3  \\ \hline
0.16 -- 0.18 &   2.02 &  16.6  &  0.1  &  0.5  &  15.8  \\ \hline
\multicolumn{6}{|c|}{  } \\ \hline
0.18 -- 0.20 &   0.41 &  12.92  &  0.09  &  0.39  &  12.61  \\ \hline
0.20 -- 0.22 &   0.21 &  10.37  &  0.09  &  0.31  &  10.18  \\ \hline
0.22 -- 0.24 &   0.46 &   8.36  &  0.08  &  0.25  &   8.15  \\ \hline
0.24 -- 0.26 &   0.33 &   6.72  &  0.07  &  0.20  &   6.80  \\ \hline
0.26 -- 0.28 &   0.58 &   5.67  &  0.06  &  0.17  &   5.51  \\ \hline
0.28 -- 0.30 &   0.11 &   4.61  &  0.06  &  0.14  &   4.64  \\ \hline
0.30 -- 0.32 &   0.32 &   3.85  &  0.05  &  0.12  &   3.90  \\ \hline
0.32 -- 0.34 &   0.15 &   3.19  &  0.05  &  0.10  &   3.21  \\ \hline
0.34 -- 0.36 &   0.15 &   2.70  &  0.04  &  0.08  &   2.66  \\ \hline
0.36 -- 0.40 &   3.89 &   2.09  &  0.03  &  0.06  &   2.21  \\ \hline
0.40 -- 0.44 &   0.04 &   1.50  &  0.02  &  0.05  &   1.48  \\ \hline
0.44 -- 0.48 &   0.19 &   1.08  &  0.02  &  0.03  &   1.09  \\ \hline
0.48 -- 0.52 &   0.05 &  0.770  & 0.016  & 0.023  &  0.761  \\ \hline
0.52 -- 0.56 &   0.50 &  0.561  & 0.014  & 0.017  &  0.570  \\ \hline
0.56 -- 0.60 &   2.31 &  0.396  & 0.011  & 0.012  &  0.376  \\ \hline
0.60 -- 0.66 &   0.66 &  0.268  & 0.007  & 0.008  &  0.260  \\ \hline
0.66 -- 0.72 &   1.97 &  0.160  & 0.006  & 0.005  &  0.167  \\ \hline
0.72 -- 0.78 &   3.53 &  0.096  & 0.004  & 0.003  &  0.088  \\ \hline
\multicolumn{6}{|c|}{  } \\ \hline
0.78 -- 0.84 &  15.69 &  0.045  & 0.002  & 0.002  &  0.055  \\ \hline
0.84 -- 0.90 &   5.72 &  0.023  & 0.002  & 0.002  &  0.019  \\ \hline
0.90 -- 1.00 &   7.58 &  0.0059 & 0.0005 & 0.0005 &  0.0044 \\ \hline
\end{tabular}
\end{center}
\caption{       The inclusive hadron $x$
spectrum as measured by DELPHI as well as the statistical
and systematic errors.
  The prediction from the  exact QCD Matrix Element calculation
followed by
string fragmentation is shown under the label QCD+SF and the
$\chi^2$ of each bin is shown in the second column.
The centre of mass energy is 91.2 GeV and the
overall normalisation factor from the fit
is 0.995 (not included in the data column).
Only the data  between the empty rows was used
for the determination of $\as$. }
\label{t1}
\end{table}
\newpage
\begin{table}
\begin{center}
\begin{tabular}{|l|c|c|}
\hline
 &$ f=0.01$ &$ f=1.0$ \\
\hline
$\lms$&$193^{+20}_{-11}$\ MeV &$ 269^{+17}_{-14}$\ MeV \\
$a$&$0.85\pm 0.03$&$0.96\pm 0.03 $ \\
$\epsilon_b$&$ 0.009\pm 0.002$&$0.008\pm 0.002 $ \\
\hline
$\chi^2/{\rm data points}    $&$ 1.02 $&$0.99 $ \\
corr. $a-\lms$          &$-0.07 $&$-0.06$ \\
corr. $\epsilon_b -\lms$&$ -0.36   $&$-0.31$ \\
corr. $\epsilon_b -a   $&$ -0.07   $&$-0.22$ \\
\hline
\end{tabular}
\end{center}
\caption{ Results of the fit  to all data (71 data points for
$Q^2 > 29^2 $ GeV$^2$ and $0.18<x<0.8$)
for two renormalisation scales
($f=\mu^2/s=0.01$ and 1.0, respectively).   }
\label{t0}
\end{table}
%
\begin{table}
\begin{center}
\begin{tabular}{|c|c|c|c|c|}
\hline
x-Bin &$\chi^2$ &Data &$\sigma_{exp}$&
QCD+SF \\           \hline
0.02 -- 0.03 &   1.99 & 169.3  &  2.4  & 173.7  \\ \hline
0.03 -- 0.04 &   1.37 & 143.7  &  2.7  & 146.4  \\ \hline
0.04 -- 0.05 &   3.74 & 115.5  &  1.6  & 120.3  \\ \hline
0.05 -- 0.06 &   3.14 &  93.3  &  1.5  &  96.7  \\ \hline
0.06 -- 0.08 &   2.52 &  69.2  &  1.2  &  71.3  \\ \hline
0.08 -- 0.10 &   0.00 &  49.7  &  1.1  &  49.0  \\ \hline
0.10 -- 0.12 &   0.13 &  36.3  &  0.4  &  36.1  \\ \hline
0.12 -- 0.14 &   0.00 &  28.1  &  0.4  &  27.7  \\ \hline
0.14 -- 0.16 &   0.29 &  22.4  &  0.4  &  21.7  \\ \hline
0.16 -- 0.18 &   0.08 &  18.0  &  0.3  &  17.6  \\ \hline
\multicolumn{5}{|c|}{  } \\ \hline
0.18 -- 0.20 &   0.01 &  14.38  &  0.28  &  14.18  \\ \hline
0.20 -- 0.25 &   0.06 &  10.24  &  0.16  &  10.16  \\ \hline
0.25 -- 0.30 &   0.01 &   6.43  &  0.11  &   6.35  \\ \hline
0.30 -- 0.35 &   0.02 &   4.23  &  0.10  &   4.18  \\ \hline
0.35 -- 0.40 &   2.18 &   2.72  &  0.09  &   2.79  \\ \hline
0.40 -- 0.50 &   0.16 &   1.59  &  0.04  &   1.54  \\ \hline
0.50 -- 0.60 &   3.75 &  0.782  & 0.028  &  0.725  \\ \hline
0.60 -- 0.70 &   2.12 &  0.341  & 0.023  &  0.304  \\ \hline
0.70 -- 0.80 &   5.46 &  0.162  & 0.018  &  0.119  \\ \hline
\multicolumn{5}{|c|}{  } \\ \hline
0.80 -- 1.00 &   0.10 &  0.0300 & 0.0120 &  0.0259 \\ \hline
\end{tabular}
\end{center}
\caption{As Table 1,        but for the TASSO    experiment[6].
The centre of mass energy is 35   GeV and the
overall normalisation factor from the fit is 0.984
(not included in the data column).
The fourth column
               includes both the statistical and systematic errors. }
\label{t2}
\end{table}
\newpage
\begin{table}
\vspace{-0.5cm}
\begin{center}
\begin{tabular}{|c|c|c|c|c|c|}
\hline
x-Bin &$\chi^2$ &Data &$\sigma_{stat}$&$\sigma_{sys}$&
QCD+SF \\           \hline
0.00 -- 0.01 &   0.04 &  62.0  &  2.0  & 69.8  &  76.0  \\ \hline
0.01 -- 0.02 &   7.46 & 153.3  &  1.1  &  7.8  & 174.2  \\ \hline
0.02 -- 0.03 &  12.80 & 155.8  &  1.1  &  4.9  & 173.7  \\ \hline
0.03 -- 0.04 &   4.73 & 136.5  &  1.0  &  4.1  & 146.4  \\ \hline
0.04 -- 0.05 &   5.59 & 111.6  &  0.9  &  3.3  & 120.3  \\ \hline
0.05 -- 0.06 &   8.54 &  88.3  &  0.8  &  2.7  &  96.7  \\ \hline
0.06 -- 0.07 &  12.30 &  71.6  &  0.7  &  2.1  &  79.7  \\ \hline
0.07 -- 0.08 &   2.22 &  59.9  &  0.7  &  2.0  &  63.0  \\ \hline
0.08 -- 0.09 &   1.32 &  50.9  &  0.6  &  1.7  &  53.0  \\ \hline
0.09 -- 0.10 &   0.20 &  44.1  &  0.6  &  1.5  &  45.0  \\ \hline
0.10 -- 0.12 &   0.14 &  36.4  &  0.4  &  1.3  &  36.1  \\ \hline
0.12 -- 0.14 &   0.05 &  27.7  &  0.3  &  1.1  &  27.7  \\ \hline
0.14 -- 0.16 &   0.00 &  21.6  &  0.3  &  0.9  &  21.7  \\ \hline
0.16 -- 0.18 &   0.02 &  17.6  &  0.3  &  0.7  &  17.6  \\ \hline
\multicolumn{6}{|c|}{  } \\ \hline
0.18 -- 0.20 &   0.04 &  14.21  &  0.24  &  0.62  &  14.18  \\ \hline
0.20 -- 0.22 &   0.01 &  11.41  &  0.21  &  0.48  &  11.53  \\ \hline
0.22 -- 0.24 &   0.06 &   9.49  &  0.19  &  0.41  &   9.65  \\ \hline
0.24 -- 0.26 &   0.68 &   8.06  &  0.18  &  0.33  &   7.83  \\ \hline
0.26 -- 0.28 &   0.30 &   6.33  &  0.16  &  0.25  &   6.52  \\ \hline
0.28 -- 0.30 &   0.29 &   5.67  &  0.15  &  0.24  &   5.57  \\ \hline
0.30 -- 0.32 &   0.51 &   4.84  &  0.14  &  0.20  &   4.71  \\ \hline
0.32 -- 0.34 &   0.73 &   4.04  &  0.13  &  0.16  &   3.91  \\ \hline
0.34 -- 0.36 &   0.07 &   3.49  &  0.12  &  0.13  &   3.47  \\ \hline
0.36 -- 0.40 &   0.20 &   2.65  &  0.07  &  0.09  &   2.62  \\ \hline
0.40 -- 0.44 &   0.24 &   1.91  &  0.06  &  0.07  &   1.88  \\ \hline
0.44 -- 0.48 &   0.00 &   1.36  &  0.05  &  0.04  &   1.37  \\ \hline
0.48 -- 0.52 &   0.65 &   1.01  &  0.04  &  0.03  &   1.05  \\ \hline
0.52 -- 0.56 &   0.38 &  0.738  & 0.035  & 0.022  &  0.766  \\ \hline
0.56 -- 0.60 &   2.08 &  0.496  & 0.027  & 0.015  &  0.541  \\ \hline
0.60 -- 0.66 &   0.56 &  0.331  & 0.018  & 0.010  &  0.348  \\ \hline
0.66 -- 0.72 &   0.47 &  0.197  & 0.012  & 0.006  &  0.207  \\ \hline
0.72 -- 0.78 &   0.59 &  0.108  & 0.008  & 0.005  &  0.116  \\ \hline
\multicolumn{6}{|c|}{  } \\ \hline
0.78 -- 0.84 &   0.50 &  0.047  & 0.004  & 0.003  &  0.051  \\ \hline
%
\end{tabular}
\end{center}
\caption{As Table 1, but for the CELLO    experiment[7].
The centre of mass energy is 35   GeV and the
overall normalisation factor from the fit is 1.007
(not included in the data column). }
\label{t3}
\end{table}
\newpage
\begin{table}
\begin{center}
\begin{tabular}{|c|c|c|c|c|}
\hline
x-Bin &$\chi^2$ &Data  &$\sigma_{exp}$&
QCD+SF \\           \hline
0.00 -- 0.05 &   1.71 & 115.4  &  1.8  & 115.7  \\ \hline
0.05 -- 0.10 &   3.26 &  65.1  &  1.2  &  71.6  \\ \hline
0.10 -- 0.15 &   6.29 &  31.6  &  0.6  &  30.5  \\ \hline
0.15 -- 0.20 &   6.33 &  17.5  &  0.3  &  16.9  \\ \hline
\multicolumn{5}{|c|}{  } \\ \hline
0.20 -- 0.25 &   2.60 &  10.40  &  0.21  &  10.33  \\ \hline
0.25 -- 0.30 &   0.00 &   6.29  &  0.13  &   6.58  \\ \hline
0.30 -- 0.35 &   0.07 &   4.07  &  0.09  &   4.21  \\ \hline
0.35 -- 0.40 &   0.05 &   2.76  &  0.07  &   2.90  \\ \hline
0.40 -- 0.45 &   2.01 &   1.80  &  0.06  &   1.96  \\ \hline
0.45 -- 0.50 &   5.16 &   1.18  &  0.04  &   1.32  \\ \hline
0.50 -- 0.55 &   0.37 &  0.810  & 0.039  &  0.868  \\ \hline
0.55 -- 0.60 &   1.43 &  0.515  & 0.031  &  0.574  \\ \hline
0.60 -- 0.65 &   0.23 &  0.347  & 0.023  &  0.351  \\ \hline
0.65 -- 0.70 &   2.45 &  0.227  & 0.020  &  0.269  \\ \hline
0.70 -- 0.75 &   0.43 &  0.167  & 0.020  &  0.161  \\ \hline
0.75 -- 0.80 &   1.24 &  0.104  & 0.016  &  0.090  \\ \hline
\multicolumn{5}{|c|}{  } \\ \hline
0.80 -- 0.85 &   0.11 &  0.062  & 0.013  &  0.069  \\ \hline
0.85 -- 0.90 &   2.93 &  0.025  & 0.007  &  0.039  \\ \hline
0.90 -- 0.95 &   0.76 &  0.013  & 0.005  &  0.018  \\ \hline
0.95 -- 1.00 &   1.60 &  0.012  & 0.006  &  0.005  \\ \hline
\end{tabular}
\end{center}
\caption{As Table 1, but for the MARK II  experiment[8].
The centre of mass energy is 29   GeV and the
overall normalisation factor from the fit is 1.044
(not included in the data column).
The fourth column
               includes both the statistical and systematic errors.
               }
\label{t4}
\end{table}
\newpage
\begin{table}
\begin{center}
\begin{tabular}{|c|c|c|c|c|}
\hline
x-Bin &$\chi^2$ &Data  &$\sigma_{exp}$&
QCD+SF \\           \hline
0.00 -- 0.10 &  18.93 & 134.3  &  1.0  & 152.6  \\ \hline
0.10 -- 0.20 &   1.06 &  23.5  &  0.4  &  22.9  \\ \hline
\multicolumn{5}{|c|}{  } \\ \hline
0.20 -- 0.30 &   0.05 &   7.79  &  0.22  &   7.77  \\ \hline
0.30 -- 0.40 &   0.02 &   3.17  &  0.15  &   3.16  \\ \hline
0.40 -- 0.50 &   1.35 &   1.23  &  0.09  &   1.34  \\ \hline
0.50 -- 0.60 &   2.10 &  0.532  & 0.050  &  0.605  \\ \hline
0.60 -- 0.70 &   1.42 &  0.310  & 0.038  &  0.267  \\ \hline
0.70 -- 0.80 &   1.40 &  0.124  & 0.022  &  0.099  \\ \hline
\multicolumn{5}{|c|}{  } \\ \hline
0.80 -- 0.90 &   6.25 &  0.021  & 0.006  &  0.036  \\ \hline
0.90 -- 1.00 &   5.65 &  0.0038 & 0.0018 &  0.0081 \\ \hline
\end{tabular}
\end{center}
\caption{As Table 1, but for the AMY      experiment[16].
The centre of mass energy is 54   GeV and the
overall normalisation factor from the fit is 1.004
(not included in the data column).
The fourth column
               includes both the statistical and systematic errors.
               }
\label{t5}
\end{table}
\newpage
\begin{table}
\begin{center}
\begin{tabular}{|c|c|c|c|c|c|}
\hline
x-Bin &$\chi^2$ &Data &$\sigma_{stat}$&$\sigma_{sys}$&
QCD+SF \\           \hline
0.005 -- 0.010 &  25.57 & 514.9  &  2.5  & 11.6  & 429.8  \\ \hline
0.010 -- 0.015 &   0.00 & 451.3  &  2.1  &  6.8  & 444.4  \\ \hline
0.015 -- 0.020 &  27.51 & 355.9  &  1.8  &  4.2  & 405.3  \\ \hline
0.02 -- 0.03 &   9.34 & 262.0  &  1.1  &  2.8  & 281.5  \\ \hline
0.03 -- 0.04 &   3.79 & 184.3  &  0.9  &  1.4  & 192.0  \\ \hline
0.04 -- 0.05 &   1.37 & 136.7  &  0.8  &  0.9  & 139.3  \\ \hline
0.05 -- 0.06 &   1.07 & 103.0  &  0.7  &  0.6  & 104.5  \\ \hline
0.06 -- 0.07 &   0.00 &  83.3  &  0.6  &  0.4  &  82.0  \\ \hline
0.07 -- 0.08 &   0.10 &  67.7  &  0.6  &  0.6  &  66.0  \\ \hline
0.08 -- 0.09 &   1.15 &  56.1  &  0.5  &  0.3  &  53.4  \\ \hline
0.09 -- 0.10 &   0.94 &  47.0  &  0.5  &  0.2  &  44.9  \\ \hline
0.10 -- 0.12 &   1.55 &  37.0  &  0.3  &  0.2  &  35.1  \\ \hline
0.12 -- 0.14 &   1.20 &  27.9  &  0.3  &  0.1  &  26.6  \\ \hline
0.14 -- 0.16 &   1.07 &  21.3  &  0.2  &  0.1  &  20.3  \\ \hline
0.16 -- 0.18 &   2.23 &  16.8  &  0.2  &  0.1  &  15.8  \\ \hline
\multicolumn{6}{|c|}{  } \\ \hline
0.18 -- 0.20 &   4.78 &  13.71  &  0.19  &  0.13  &  12.61  \\ \hline
0.20 -- 0.25 &   0.00 &   8.93  &  0.09  &  0.12  &   8.80  \\ \hline
0.25 -- 0.30 &   0.06 &   5.43  &  0.07  &  0.08  &   5.38  \\ \hline
0.30 -- 0.40 &   0.05 &   2.88  &  0.04  &  0.04  &   2.85  \\ \hline
0.40 -- 0.50 &   0.08 &   1.24  &  0.02  &  0.02  &   1.21  \\ \hline
0.50 -- 0.60 &   0.04 &  0.534  & 0.016  & 0.012  &  0.522  \\ \hline
0.60 -- 0.70 &   0.13 &  0.230  & 0.011  & 0.004  &  0.230  \\ \hline
0.70 -- 0.80 &   1.44 &  0.090  & 0.006  & 0.002  &  0.096  \\ \hline
\multicolumn{6}{|c|}{  } \\ \hline
\end{tabular}
\end{center}
\caption{As Table 1, but for the ALEPH    experiment[25].
The centre of mass energy is 91.2 GeV and the
overall normalisation factor from the fit is 0.984
(not included in the data column). }
\label{t6}
\end{table}
\newpage
\begin{table}
\begin{center}
\begin{tabular}{|l|c|}
\hline
Source        &Error on $\lms$  \\
\hline
Errors from Fit     &   $^{+20}_{-11}$ \ MeV\\
Combinations of experiments&   $\pm 15$ \ MeV\\
Heavy Quark Fractions      &   $\pm 10$ \ MeV\\
Fragmentation dependence   &   $\pm  9$ \ MeV\\
$x$ dependence&   $\pm 10$ \ MeV\\
Cut-off dependence&   $\pm 30$ \ MeV\\
Scale   dependence&   $\pm 40$ \ MeV\\
\hline
Total                  &  $\pm 60  $   \ MeV\\
\hline
\end{tabular}
\end{center}
\caption{Summary of systematic errors.
The total error was obtained by adding quadratically all errors. }
\label{t7}
\end{table}
\newpage


\newpage

\begin{figure}[t]
 \begin{center}
  \leavevmode
  \epsfxsize=16cm
  \epsfysize=16cm
  \epsffile{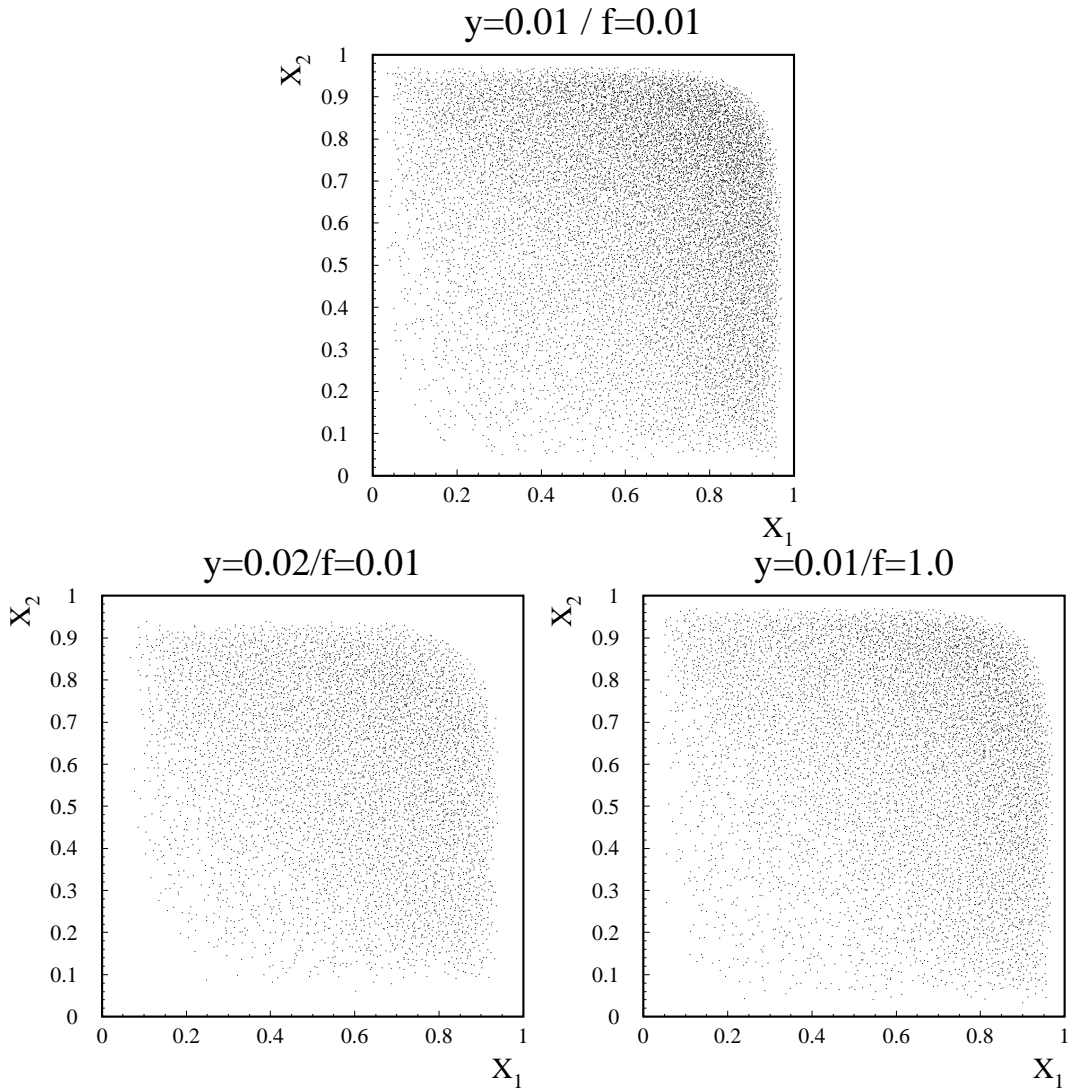}
 \end{center}
 \caption{The phase space distributions of 4-jet events at an energy
 of 91 GeV for various values of the cut-off parameter
 $y$ and the renormalisation scale factor $f$.}
 \label{f6}
\end{figure}

\end{document}